\documentclass[aps,twocolumn,amsfonts,nofootinbib]{revtex4}

\usepackage{amsmath}
\usepackage{graphicx}

\def\be{\begin{equation}}
\def\ee{\end{equation}}
\def\bea{\begin{eqnarray}}
\def\eea{\end{eqnarray}}
\def\rx {\rho_{\text{eff}}}
\def\wx {w_{\text{eff}}}
\def\orc{\Omega_{r_c}}
\def\om{\Omega_{\text{m}}}
\def\ol{\Omega_\Lambda}
\def\oq{\Omega_{\text{q}}}
\def\rq {\rho_{\text{q}}}
\def\wq {w_{\text{q}}}

\begin{document}

\title{Crossing the phantom divide without phantom matter}

\author{Luis P. Chimento$^1$, Ruth Lazkoz$^2$, Roy Maartens$^3$,
Israel Quiros$^{3,4}$}

\affiliation{\vspace*{0.2cm} $^1$Departamento de F\'isica,
Universidad de Buenos Aires, 1428 Buenos Aires, Argentina\\
$^2$Fisika Teorikoa, Euskal Herriko Unibertsitatea, 48080 Bilbao,
Spain\\
$^3$Institute of Cosmology \& Gravitation, University of
Portsmouth, Portsmouth PO1 2EG, UK\\
$^4$Departamento de F\'isica, Universidad Central de Las Villas,
Santa Clara CP 54830, Cuba}

%\email{chimento@df.uba.ar}\email{ruth.lazkoz@ehu.es}
%\email{roy.maartens@port.ac.uk}\email{israel@uclv.edu.cu}

\date{\today}

\begin{abstract}

A class of braneworld models can lead to phantom-like acceleration
of the late universe, but without the need for any phantom matter.
In the simplest models, the universe contains only cold dark
matter and a cosmological constant. We generalize these models by
introducing a quintessence field. The new feature in our models is
that quintessence leads to a crossing of the phantom divide,
$w=-1$. This is a purely gravitational effect, and there is no
phantom instability. Furthermore, the Hubble parameter is always
decreasing, and there is no big rip singularity in the future.

\end{abstract}

%\pacs{PACS numbers: 04.20.Jb, 04.20.Dw, 98.80.-k, 98.80.Es,
%95.30.Sf, 95.35.+d}

\maketitle

%%%%%%%%%%%%%%%%%%%%%%%%%%%%%%%%%%%%%%%%%%%%%%%%%%%%%%%%%%%%%%%%%%%%%%%%

\section{Introduction}

Observations of supernovae redshifts, cosmic microwave background
anisotropies and the large-scale structure provide increasingly
strong evidence that the late-time universe is accelerating. If
general relativity is a correct description of large-scale
gravity, then the acceleration typically originates from a dark
energy field with $w \equiv p/\rho<-{1\over3}$. The simplest model
is LCDM, where the dark energy is the vacuum energy ($w=-1$). It
gives a very good fit to the data~\cite{Spergel:2006hy}, but with
an unnaturally small and fine-tuned value of $\Lambda$.
Quintessence scalar fields, with $w>-1$, produce more general
dynamical behaviour, but they do not improve the fit to the data,
and also do not solve the theoretical problems faced by LCDM.
``Phantom" scalar fields, with $w<-1$, have the same theoretical
problems as quintessence, but in addition, classically they have
the unphysical behaviour that the energy density increases with
the expansion of the universe, and they also lead to instability
of the quantum vacuum.

Current data does not rule out a phantom value, $w<-1$. The WMAP
3-year data, in combination with large-scale structure and SN
data, allows for $w<-1$; in a flat general relativistic
model~\cite{Seljak:2006bg},
 \be
w=-1.04 \pm .06\,.
 \ee
Models with $w<-1$ but without the problem of quantum instability
are therefore still an interesting possibility. An effective $w$
such that $w<-1$ can occur in modified gravity theories, without
any phantom matter that renders the quantum vacuum unstable.
Instead, gravity itself produces phantom-like acceleration. Sahni
and Shtanov~\cite{Sahni:2002dx} showed that a class of braneworld
models exhibited effective phantom behaviour. In these models, the
4-dimensional brane universe contains only matter and a
cosmological constant $\Lambda$. A 5-dimensional gravitational
effect screens $\Lambda$, leading to phantom dynamics.

These models are a variant of the Dvali-Gabadadze-Porrati (DGP)
braneworld model, generalized to cosmology by
Deffayet~\cite{Dvali:2000rv}. The standard DGP model is a
self-accelerating model without any form of dark energy, and the
effective $w$ is always non-phantom. However, there is another
branch of DGP models, with a different embedding of the 4D brane
universe in the 5D bulk spacetime. The self-accelerating DGP model
is the (+) branch. The DGP$(-)$ model is very different. It does
not self-accelerate, but requires dark energy on the brane. It
experiences 5D gravitational modifications to its dynamics, which
effectively screen the dark energy. At late times, as gravity
leaks off the 4D brane, the dynamics deviates from general
relativity. The transition from 4D to 5D behaviour is governed by
a crossover scale $r_c$, the same as in the DGP(+) branch.

The simplest DGP$(-)$ model has a cosmological constant $\Lambda$,
and we follow Lue and Starkman~\cite{Lue:2004za} and call this the
LDGP model. The dynamics of these models were investigated by Lue
and Starkman~\cite{Lue:2004za}, and observational constraints on
the models have been considered by Lazkoz et al.~\cite{lmm}. The
energy conservation equation for CDM remains the same as in
general relativity, but the Friedman equation is modified:
\begin{eqnarray}
&& \dot\rho+3H\rho=0\,,\label{ec} \\ && H^2+{H \over r_c}=
{1\over3}(\rho +\Lambda)\,. \label{f}
\end{eqnarray}
[The DGP($\pm$) branches have $\mp H/r_c$ in Eq.~(\ref{f}).] These
equations imply
 \be
\dot H=-{\rho \over 2} \left[ 1 - {1\over \sqrt{1+ 4
r_c^2(\rho+\Lambda)/3}} \right]. \label{r}
 \ee
Equation~(\ref{f}) shows that at early times, i.e., for $H \gg
r_c^{-1}$, the general relativistic Friedman equation is
recovered, but at late times, the $H/r_c$ term is important and
the Friedman equation is nonstandard.

At late times, gravity leakage screens the cosmological constant,
leading to an effective dark energy~\cite{Sahni:2002dx,Lue:2004za}
 \be \label{rx}
\rx=\Lambda-3{H \over r_c}\,,
 \ee
where $\rx$ and $\wx= p_{\text{eff}}/\rx$ are effective quantities
in a general relativistic interpretation of LDGP expansion
history, i.e., they describe the equivalent general relativity
model:
 \bea
&& H^2= {1\over 3}(\rho+\rx)\,,\label{f2}\\ &&
\dot{\rho}_{\text{eff}}+3H(1+\wx)\rx=0\,. \label{rp}
 \eea
It follows that
 \be \label{wx}
1+\wx={\dot H \over r_cH\rx}\,,
 \ee
and since $\dot H< 0$ by Eq.~(\ref{r}), we have effective phantom
behaviour, $\wx<-1$, provided that $\rx>0$, i.e., for $H<
r_c\Lambda/3$.

In general relativity with phantom matter, $\wx<-1$ implies that
$\dot H$ eventually becomes positive, i.e., the universe
eventually super-accelerates, which can lead to a ``big rip"
singularity. By contrast, in LDGP $\dot H$ is always negative, and
there is no big rip singularity. In LDGP however, the phantom
divide $\wx=-1$ cannot be crossed: $\wx$ is always less than $-1$
for $\rx>0$. At higher redshifts, when $H>r_c\Lambda/3$, we have
$\wx>-1$, but $\wx$ never passes through $-1$, so there is no
crossing of $\wx=-1$. The effective phantom picture only holds for
$\rx>0$.

In this paper we generalize the LDGP model by introducing a
quintessence field with $\wq=$const ($\geq -1$),
%i.e., with an exponential potential,
% \be
%V(\phi)= V_0\exp(\alpha \phi)\,,
% \ee
sometimes called ``quiessence". We will call this the QDGP model.
We will show that crossing of the phantom divide does occur in
QDGP models, unlike the LDGP limit. The interesting fact is that
in general relativity with a dark energy component, phantom
crossing cannot occur with a single field~\cite{Chimento:2006xu}
(unless non-minimal coupling between dark matter and dark energy
is allowed~\cite{Curbelo:2005dh}). In the QDGP model, crossing is
possible without resorting to a complicated model with multiple
fields or additional non-gravitational interactions.

\section{The QDGP Model}

\begin{figure}[!h]
\begin{center}
\includegraphics[width=8cm]{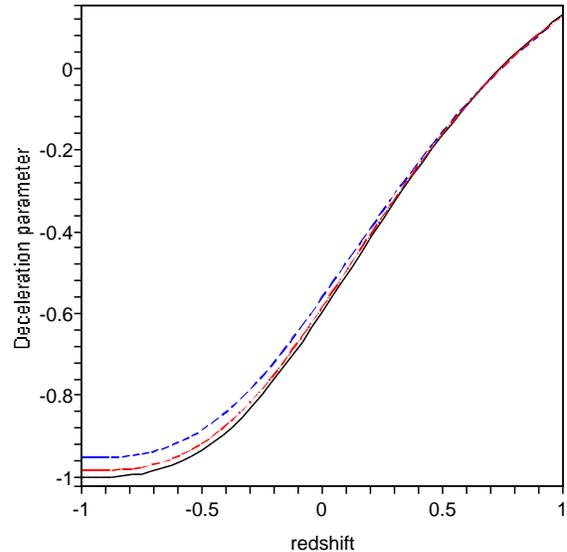}
\caption{ The deceleration parameter $q=-(1+\dot E/H_0E^2)$
against redshift, for $(\om,\oq,\wq)=(0.28,0.8,-1)$ [solid (black)
curve], $(0.28,0.8,-0.99)$ [dotted (red) curve], and
$(0.28,0.8,-0.95)$ [dashed (blue) curve]. }\label{qdgp2}
\end{center}
\end{figure}

The Friedman equation is Eq.~(\ref{f2}), with
 \bea
\rx &=& \rq-3\frac{ H}{r_c}\,.\label{rhoeff}
 \eea
Since $\wq$ is constant, the conservation equations give
 \be
\rho=\rho_{0}(1+z)^{3}\,,~~\rq=\rho_{\text{q}0}(1+z)^{3(1+\wq)}\,.
\label{densities}
 \ee
We define dimensionless density parameters,
 \be
\om=\frac{\rho_{0}}{3H_0^2}\,,~~\oq=\frac{\rho_{\text{q}0}}
{3H_0^2}\,,~~\orc=\frac{1}{4 r_c^2 H_0^2}\,.\label{phenomparam}
 \ee
Then Eqs.~(\ref{f2}) and (\ref{rhoeff}) give
 \bea
\!\! E(z)\!\equiv \! \frac{H}{H_0}\! &=&\! \sqrt{\orc+\om
(1+z)^3+\oq(1+z)^{3(1+\wq)}}
\nonumber\\
&&~{}-\sqrt{\orc}\,,\label{Ea}
 \eea
so that
 \be
2 \sqrt{\orc}=\om+\oq-1 \geq 0\,.\label{link}
 \ee
Thus the flat QDGP model mimics a closed GR quiessence model in
the $(\om,\oq)$ plane, as in the LDGP case~\cite{lmm}.

The evolution of the Hubble parameter is given by
 \be
{\dot E \over H_0} =-\frac{3E (1+z)^3\left[\om +(1+\wq) \oq
(1+z)^{3\wq}\!\right]}{2\left(E+\sqrt{\orc}\right)}.
 \ee
This equation shows that $\dot E<0$ for all $z$, as in the LDGP
case, and that $\dot E \to 0$ as $z\to -1$ (equivalently, as
$a\to\infty$), for all $\wq\geq -1$. This behaviour is illustrated
in Fig.~\ref{qdgp2}. Thus there is no super-acceleration, and no
big rip singularity in QDGP models. The purely gravitational
phantom behaviour does not produce the pathologies associated with
phantom matter in GR.

Since $\dot E$ is always negative, the deceleration parameter,
$q=-(1+\dot E/H_0E^2)$, can be never less than $-1$, as seen in
Fig.~\ref{qdgp2}. Note that the asymptotic future value of $q$
depends on $\wq$:
 \be
q\to 3(1+\wq)-1 ~~\text{as} ~~ z\to -1\,.
 \ee
The transition redshift $z_{\text{a}}$ at which decelerated
expansion turns into accelerated expansion, is defined by $q=0$,
i.e.,
 \be
\dot E(z_{\text{a}})=-H_0E(z_{\text{a}})^2\,.
 \ee
For the parameter values in Fig.~\ref{qdgp2}, $z_{\text{a}} \sim
0.7$.

The QDGP and LDGP models show an important difference in the
asymptotic behaviour of the Hubble rate:
 \be
E \to \left\{ \begin{array}{cll} 0\,, & $~~$& \wq>-1\,, \\
\sqrt{\orc+\ol}-\sqrt{\orc}>0\,, & & \wq=-1\,. \end{array} \right.
 \ee
In LDGP models, the de Sitter solution is a stable attractor. This
arises because the vacuum energy does not redshift. By contrast,
in QDGP models we have $H\to 0$, since the quintessence is
redshifting away.

\begin{figure}[t]
\begin{center}
\includegraphics[width=8cm]{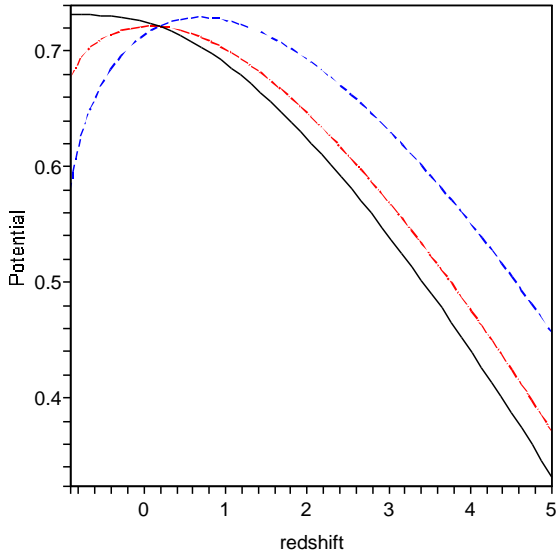}
\caption{ The equivalent GR phantom potential $V_{\text{eff}}(z)$
for the parameter values in Fig.~\ref{qdgp2}.}\label{qdgp4}
\end{center}
\end{figure}

The quantities $\rx$ and $\wx$ describe a phantom scalar field
$\phi_{\text{eff}}$ in GR which has the equivalent expansion
history to the QDGP model. We can find the self-interaction
potential $V_{\text{eff}}(\phi_{\text{eff}})$ of the equivalent GR
phantom field by using the well-known result that a scalar field
and its potential may be determined for any given expansion
history as follows~\cite{Ellis:1990ws}:
 \bea
\frac{V_{\text{eff}}(z)}{3 H_0^2}=E^2-\om (1+z)^3- \frac{d[E^2-\om
(1+z)^3]}{2\;\; d\ln(1+z)^3}, \label{eqv}\\
{\phi_{\text{eff}}(z)\over \sqrt{3}}=-\int\frac{dz}{(1+z) E}\left[
\frac{d[E^2-\om (1+z)^3]} {d\ln(1+z)^3}\right]^{1/2}.\label{eqp}
 \eea
In principle, these equations lead to
$V_{\text{eff}}=V_{\text{eff}}(\phi_{\text{eff}})$, but in
practice the inversion cannot be performed analytically because of
the integral in Eq.~(\ref{eqp}). In Fig.~\ref{qdgp4} we show
$V_{\text{eff}}(z)$. For the LDGP model, $V_{\text{eff}}$
approaches a constant nonzero value as $a\to\infty$. By contrast,
in the QDGP model the potential vanishes asymptotically.

\section{Dynamical System Analysis}

In order to write the cosmological equations in the form of an
autonomous system, we define the following expansion normalized
variables:
 \be
x=\frac{\sqrt{\om}}{a^{3/2}E}\,,~ y={\frac{\sqrt{\oq
}}{a^{3(1+\wq)/2} E}}\,,~ v={\frac{\sqrt\orc}{E}}\,,
\label{variables}
 \ee
so that Eq.~(\ref{Ea}) becomes
 \be
2v=x^2+y^2-1\,.\label{constraint}
 \ee
This constraint means that the phase space is defined by
$x^2+y^2\geq 1$, since $v\geq0$, and by $x\geq1, y\geq1$.
Introducing the new time variable $\tau=\ln a$, and eliminating
$v$ and $E$, we obtain the autonomous system
 \bea
x'&=&\frac{3x\left[x^2+(1+2\wq)y^2-1\right]}{2(x^2+y^2+1)},
\label{ds1}\\
y'&=& \frac{3y
\left[2x^2+(1+\wq)(y^2-x^2-1)\right]}{2(x^2+y^2+1)}.\label{ds2}
 \eea
The analysis depends on whether $\wq>-1$ or $\wq=-1$.

\begin{table}[b!]
\caption[crit]{Fixed points  for $0>\wq>-1$. }
\begin{tabular}
{@{\hspace{2pt}}c@{\hspace{8pt}}
c@{\hspace{8pt}}c@{\hspace{8pt}}c@{\hspace{8pt}}%c@{\hspace{8pt}}
c@{\hspace{4pt}}} \hline
\hline\\[-0.3cm]
Point & $(x,y)$ & Eigenvalues & Character \\[0.1cm]
\hline\\[-0.2cm]
$M$ & $(1,0)$ & $\left(-\displaystyle
\frac{3\wq}{2},\displaystyle\frac{3}{2}\right)$ &repellor\\[0.6cm]
%%%%%
$Q$& $(0,1)$ & $\left(\displaystyle\frac{3 \wq}{2},
\displaystyle\frac{3 (1+\wq)}{2}\right)$ &saddle\\[0.4cm]
\hline \hline
\end{tabular}\label{tab1}
\end{table}

\begin{table}[t!]
\caption[crit]{Fixed points  for $\wq=-1$. }
\begin{tabular}
{@{\hspace{2pt}}c@{\hspace{8pt}}c@{\hspace{8pt}}c@{\hspace{8pt}}
c@{\hspace{8pt}}c@{\hspace{4pt}}} \hline
\hline\\[-0.3cm]
Point &$y$&Eigenvalues&Character\\[0.1cm]
\hline\\[-0.2cm]
$M$ &  $0$
& $\displaystyle\frac{3}{2}$ &repellor\\[0.6cm]
%%%%%
$dS$& $\sqrt{\orc/\ol}+\sqrt{1+\orc/\ol}$  &$-3
$&attractor\\[0.4cm]
\hline \hline
\end{tabular}\label{tab2}
\end{table}

\begin{figure}[t!]
\begin{center}
\includegraphics[width=7cm]{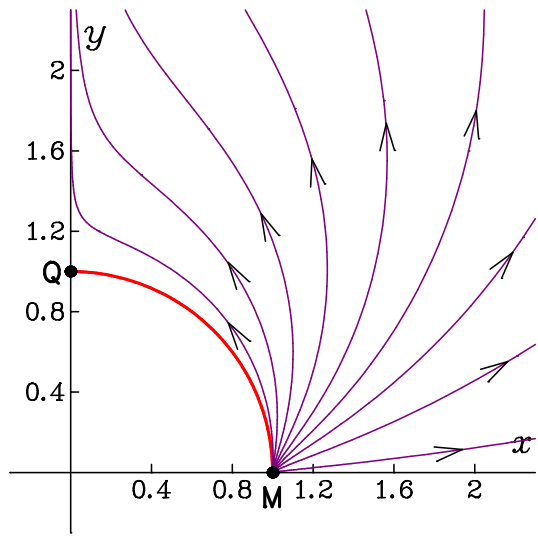}
\caption{Phase plane trajectories for $\wq=-0.8$. } \label{quin}
\end{center}\end{figure}

\begin{figure}[t!]
\begin{center}
\includegraphics[width=6.5cm,height=2.2cm]{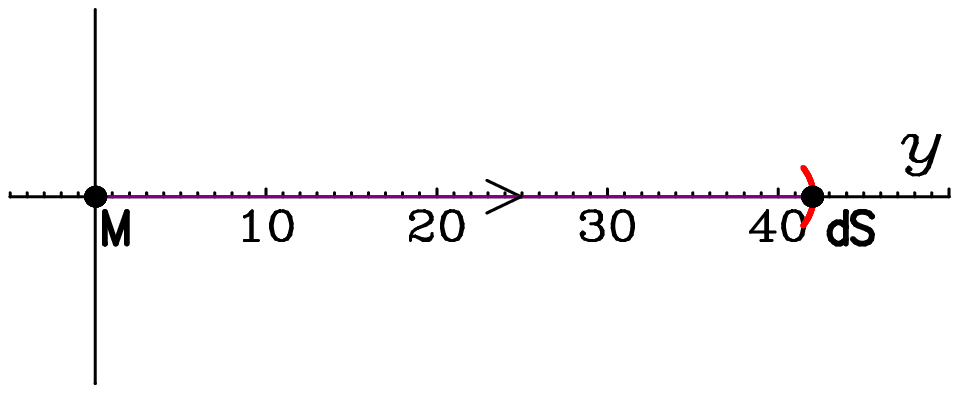}
\caption{Phase line trajectories for $\wq=-1$,
$\sqrt{\orc/\ol}=21$.} \label{lambda}
\end{center}
\end{figure}

For $\wq> -1$, the phase space is 2D. The fixed points and their
stability are summarized in Table~\ref{tab1}. In this case there
are two isolated fixed points located at finite values of $x$ and
$y$. Combining this fact with a simple examination of the $y'$
equation and the open topology of the phase space, we can draw
fairly general conclusions about the properties of the fixed
points of the system even before doing the linear analysis. Since
$y>0$ and $\wq>-1$, we have $y'>0$, so that the fixed points
cannot be stable. In addition, the fixed points cannot be unstable
spirals or centres, so they can only be unstable nodes
(attractors) or saddle points. The linear analysis confirms these
findings. The fixed points $M$ (early time) and $Q$ (late time)
are matter- and quintessence-dominated solutions respectively. The
phase plane trajectories are illustrated in Fig.~\ref{quin}.

For the LDGP limit, $\wq=-1$, we have $y\propto v$, leading to a
1D phase space, and the dynamical equations reduce to
 \be
y'=\frac{3y\left(1-y^2+2\sqrt{\orc/\ol}\,y\right)}{2\left(
1+\sqrt{\orc/\ol}\, y\right)}.
 \ee
There are also two isolated fixed points at finite values of $y$
(see Table~\ref{tab2} and Fig.~\ref{lambda}). It can be seen that
$y'>0$ as well, but since the phase space is only a line segment,
and the fixed points are just its end points, it follows that one
fixed point will necessarily be unstable (the matter-dominated
point $M$), and the other will be forced to be stable (the de
Sitter attractor $dS$).

\section{Crossing the Phantom Divide}

Perhaps the most distinctive feature of the QDGP model is the
possibility of crossing the phantom divide $w=-1$. This occurs
with non-phantom quintessence, via the infrared gravitational
modifications to GR. In the standard GR model, this can be
realized only by introducing additional complicated features like
multiple dark energy fields or interactions between dark energy
and CDM.

\begin{figure}[!b]
\begin{center}
\includegraphics[width=8cm]{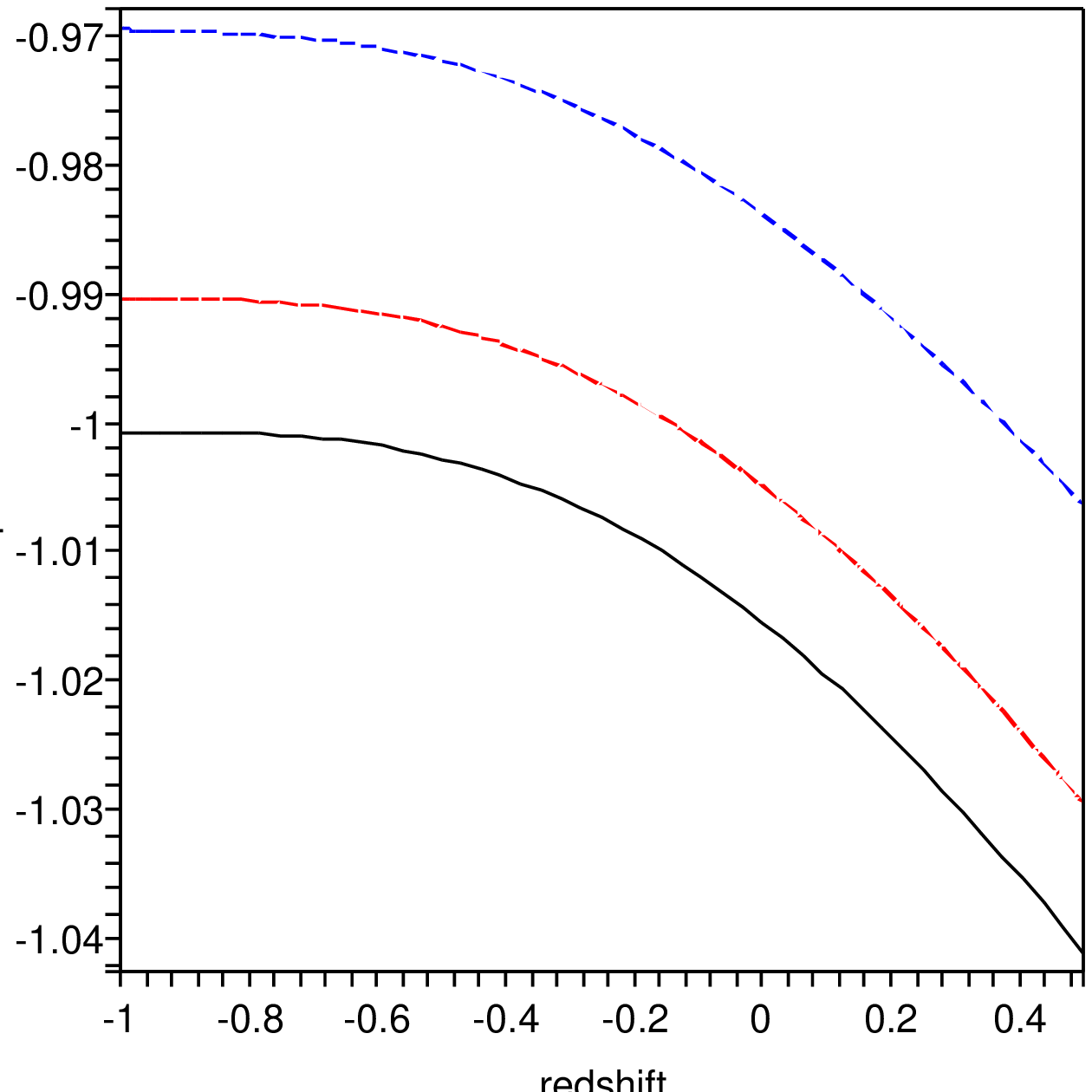}
\caption{ The effective dark energy equation of state $\wx(z)$ for
the same parameter values as in Fig.~\ref{qdgp2}.}\label{qdgp1}
\end{center}
\end{figure}

Equation~(\ref{rp}) defines the effective equation of state, which
leads to
 \bea
&&\!\!\!\!\!\!\!1+\wx(z) =\nonumber\\
&& \!\!\!\!\!\!\!\frac{(1+\wq)\oq E(z) (1+z)^{3\wq+3}-\sqrt{\orc}
\om (1+z)^3}{\left[E(z)+\sqrt{\orc}\, \right] \left[\oq
(1+z)^{3\wq+3 }-2 \sqrt{\orc} E(z)\right]}\!.\label{effeos}
 \eea

For the LDGP model, i.e., the $\wq=-1$ limit of QDGP, $\wx$ is
always more negative than $-1$, provided that $\rx>0$. When
$\rx=0$, at some redshift $z_*$, we have $\wx=-\infty$. For
$z>z_*$, we have $\wx>-1$, with $\wx\to+\infty$ as $z\to z_*$.
Thus $\wx$ never passes through $-1$.

For $\wq>-1$, the nonzero term $(1+\wq)\oq E(z)(1+z)^{3\wq+3}$ in
Eq.~(\ref{effeos}) allows for $\wx$ to pass through $-1$. This is
illustrated in Fig.~\ref{qdgp1}. For $\wq>-1$, smooth crossing of
the phantom divide occurs at a redshift $z_{\text{c}}$ that
depends on the values of the free parameters $\om$, $\oq$ and
$\wq$. By Eq.~(\ref{effeos}), $z_{\text{c}}$ is given by
 \be
(1+z_{\text{c}})^{3\wq} E(z_{\text{c}}) =
\frac{\om\sqrt{\orc}}{(1+\wq)\oq}\,.\label{crossz}
 \ee
Note that for LDGP ($\wq=-1$), the solution is $z_{\text{c}}=-1$,
i.e., crossing occurs at $a=\infty$.

At a redshift $z_*>z_{\text{c}}$, we have $\rx(z_*)=0$, so that
the effective phantom GR picture of QDGP breaks down:
 \be
z_*=\left(\frac{4 \orc\om}{\oq^2}\right)^{1/3(21+\wq)}-1\,.
 \ee
As in the LDGP case, we have $\wx\to \mp\infty$ as $z\to
z_*^{\mp}$.

\section{Conclusions}

The DGP$(-)$ class of braneworld models can lead to phantom-like
behaviour of the effective dark energy, but without the need for
any phantom matter, as pointed out by Sahni and
Shtanov~\cite{Sahni:2002dx}. In the simplest LDGP model, the
universe contains only cold dark matter and a cosmological
constant. We generalized this by introducing a quintessence field,
to define the QDGP model. QDGP has the same effective phantom
behaviour as LDGP. The Hubble parameter is always decreasing, and
there is no big rip singularity in the future. The key new feature
of QDGP is crossing of the phantom divide, $w=-1$, which is not
possible in the LDGP case.

The avoidance of any big rip is due to $H,\dot H \to 0$ as $a\to
\infty$. This asymptotic behaviour reflects the fact that the
phantom effects never dominate -- unlike the case of a phantom
scalar field in GR. The total equation of state parameter, defined
by $w_{\text{tot}}=p_{\text{tot}}/
\rho_{\text{tot}}=w_{\text{eff}}\rx/(\rho+\rx)$, follows from
Eqs.~(\ref{Ea}) and (\ref{effeos}),
 \bea
&& 1+w_{\text{tot}}(z)= \nonumber \\&&~{\om (1+z)^3+(1+\wq) \oq
(1+z)^{3(1+\wq)} \over E(z)\left[
\sqrt{\orc}+E(z)\right]}\,.\label{wt}
 \eea
This shows that $w_{\text{eff}}(z)\geq -1$.

~\\{\bf Acknowledgements}\\
We thank Mariam Bouhmadi-Lopez for useful discussions. LC, RL and
IQ thank the ICG, Portsmouth for hospitality while this work was
initiated. The visit of IQ was supported by the Royal Society.
This work has been carried out with the partial support of the
Spanish Ministry of Science and Education through research grants
FIS2004-0374-E. RL is supported by the Spanish Ministry of Science
and Education through the RyC program, and research grants
FIS2004-01626 and FIS2005-01181.

\end{document}